\documentclass[prl,twocolumn,showpacs]{revtex4}
\usepackage{graphicx}

\begin{document}
\newcommand{\ltwid}{\mathrel{\raise.3ex\hbox{$<$\kern-.75em\lower1ex\hbox{$\sim$}}}}
\newcommand{\gtwid}{\mathrel{\raise.3ex\hbox{$>$\kern-.75em\lower1ex\hbox{$\sim$}}}}

\title{Stripes on a 6-Leg Hubbard Ladder}

\author{Steven R.~White}

\affiliation{Department of Physics\\
University of California, Irvine, CA 92697}

\author{D.J.~Scalapino}

\affiliation{Department of Physics\\
University of California, Santa Barbara, CA 93106-9530}

\date{\today}

\begin{abstract}

While DMRG calculations find stripes on doped n-leg t-J ladders,
little is known about the possible formation of stripes on n-leg Hubbard
ladders.  Here we report results for a 7$\times 6$ Hubbard model with 4
holes.  We find that a stripe forms for values of $U/t$ ranging
from 6 to 20. For $U/t \sim 3-4$, the system exhibits the
domain wall feature of a stripe, but the hole density is very
broadened.

\end{abstract}
\pacs{PACS }

\maketitle


The nature of the ground state of the two dimensional t-J model
continues to be controversial, even if one restricts one's
attention to numerical simulations. 
Density matrix renormalization group (DMRG) calculations\cite{dmrg} for n-leg t-J
ladders (with n ranging from 2 to 8) with J/t in the physical region of 
interest for the cuprates, find 
that stripes are formed when the ladders are doped with
holes\cite{whitescalapino}. In contrast, the Green's function
Monte Carlo (GFMC) results of Sorella, et al., on square,
periodic t-J lattices find a $d_{x^2-y^2}$-wave superconducting
ground state\cite{sorella}. The aspect ratio of an n-leg lattice
and the open boundary conditions used in the DMRG calculations 
may be sufficient to favor a striped state. Alternatively, the
choice of the guiding trial wave function for the GFMC may bias
the system towards a superconducting state\cite{leecomment}.
It is clear that the t-J model is a delicately balanced system.
Thus it is interesting to go back one step and study the
underlying Hubbard model in which the onsite Coulomb interaction
U is a parameter.

Much less is known about the nature of the ground state of the
2D Hubbard model, which is computationally more demanding. 
Previous DMRG calculations\cite{bonca} on 3-leg Hubbard ladders found that
diagonal three-hole stripes with a linear filling density of unity formed when U/t was
greater than 5.
Early mean-field Hartree-Fock calculations for the 2D Hubbard
model found that vertical stripes were favored for U/t less than of order
4 and diagonal stripes formed at larger values of U/t \cite{schulz}.
These Hartree-Fock
stripes have a linear filling density of one and their width is set by a
coherence length $\xi_0\sim t/\Delta_0$ with $\Delta_0$ the gap of the
half-filled system. 
As U increases, $\xi_0$ decreases, approaching a lattice
spacing at larger values of U/t.  Eventually, when U exceeds twice the
bandwidth, the mean-field stripes disappear\cite{inui}. Note, however, that
the Hartree-Fock treatments do not include pairing and that their
energy per hole is of order $1 t$ too high.
Here, we extend the DMRG study of stripe formation
in the Hubbard model and in particular examine the question of whether
stripes form on a 6-leg Hubbard ladder and how their structure depends upon
U/t.  Our results represent the first reliable ground state results for a
Hubbard cluster large enough to exhibit an unambiguous stripe.
From this single cluster we are not able to make conclusions
about the infinite 2D lattice; on the other hand, our results
strongly suggest that stripes are low-lying states in the
Hubbard model, as they are in the t-J model.

The cluster geometry was chosen with some care. Stripe-like
behavior involving only two holes can be suggestive\cite{prelovsek,eder}, 
but may
only represent a feature of hole pairs. One would like at least four
holes in a stripe, which makes the system size beyond the
current reach of exact diagonalization.
Periodic boundary conditions, when the dimensions of the cluster
are even, {\it frustrate} a single stripe\cite{inui,eder}. Making one of the
dimensions odd tends to {\it force} a single stripe. We choose a 
7$\times$6 cluster, with cylindrical boundary conditions,
periodic in the $y$ direction and open in the $x$ direction, and
with four holes. Our
previous calculations on the t-J model show that cylindrical
L$\times$6 systems have ground states with four-hole stripes
wrapped around the cylinder.  The open boundaries neither
frustrate nor force the stripes, although if they form, the
boundaries serve to pin them. One may worry that Friedel oscillations 
in the charge density, induced by the open boundaries, mimic stripes,
but we find, for a number of reasons, that our stripes in the
t-J model are inconsistent with the Friedel oscillation
scenario\cite{friedel}. Sorella et al. compared GFMC and DMRG on a 6$\times$6
cylindrical t-J cluster with 6 holes\cite{sorella}; however, we
expect 6 holes to split into a stripe and a pair, with resonance
between them, obscuring any obvious signs of a stripe in the
charge or spin densities. (A stripe having an odd number of holes
would be frustrated in the $y$ direction.)

Here we begin by looking at the t-J model and then turn to the
Hubbard model.  
A typical result for a 17$\times 6$ t-J lattice with J/t $=$ 0.35 and
12 holes, with cylindrical boundary conditions, is shown in Fig.\,1.  
A maximum of $m=5000$ states were kept, for a discarded
weight of $7\times10^{-5}$.
One can see in Figure 1 that three charged stripes
have formed and the spins form $\pi$-phase-shifted regions between the
stripes.  For an L$\times$6 ladder, the linear charge density
along a stripe is 2/3. Studies of longer stripes find that the preferred
density on a long stripe is 1/2. 
Competing with the four-hole stripe, but higher in
energy per hole, is a single pair,
which also exhibits some stripe-like features.
As seen in Fig.~1, there are two types of four-hole stripes:
the stripes on the ends are (mostly) bond-centered, and the stripe in the middle is
site-centered.  The energy per unit length of these two configurations is
extremely close.  The charge and spin density profiles along the
x-direction are plotted in a more conventional fashion in Fig. 1(b).  
Since later we will be considering a 7$\times$6 ladder for the Hubbard model,
also shown is the local hole density on a 7$\times$6, t-J
ladder with 4 holes and J/t $=$ 0.35. One can see that the open
boundaries on the small system mimic the presence of the other
stripes on the larger system.

\begin{figure}
\includegraphics[width=6cm]{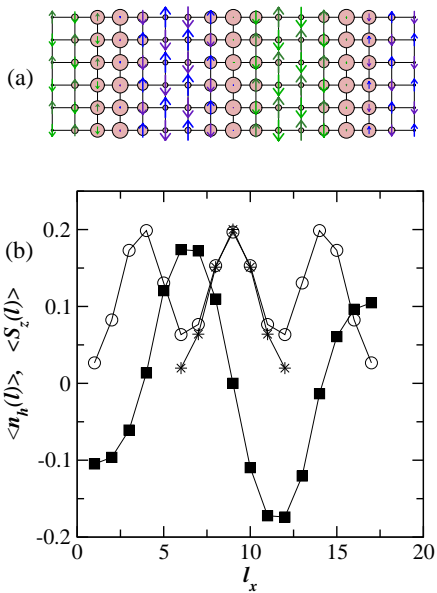}
\caption{(a) Charge and spin distribution on a doped $17\times6$, $t$-$J$ lattice
with 12 holes and $J/t = 0.35$. All the lattices that we will disucss have
periodic boundary conditions in the vertical 6-site $y$-direction and have
open boundary conditions in the x-direction. Here, on the $17\times6$
lattice, three charge stripes have formed with 4 holes each separated by
$\pi$-phase-shifted antiferromagnetic regions.
(b) Same results, plotted differently. The open circles show
the hole density $\langle n_h(l)\rangle$ and the squares show
the spin density $\langle S_z(l)\rangle (-1)^l$.
In addition, the stars show
results for a 7$\times$6 system with 4 holes, shifted to the
center of the system.
} 
\label{figone}
\end{figure}
   
In Fig. 1, the presence of static magnetic moments makes the $\pi$-phase shift 
in the antiferromagnetic spin density as one crosses a stripe clearly visible. 
No external magnetic fields were applied to the system, and these static 
magnetic moments are artifacts of the DMRG calculation, for which no spin symmetry
has been used.  However, we argue that the presence of these
static moments has very little effect on the validity of the results. To see this,
we show DMRG results for an undoped 17$\times$6 system in Fig.~2. The undoped
system converges very rapidly with the number of states $m$, allowing us
to study the convergence in detail. In particular, we consider 
$\langle S_z(l)\rangle$ versus the error in energy for a particular
number of states kept $m$.  The exact
ground state of this system, by the Lieb-Mattis theorem, is a spin singlet,
which is spin-rotationally invariant,
and therefore $\langle S_z(l)\rangle = 0$ for all sites $l$. However, 
as measured by the spin-spin correlation function,
this state has long-range antiferromagnetic order.
The ground state is a superposition of all the static antiferromagnetic states with
all possible orientations of the order parameter. 
The reduction in energy
due to the macroscopic superposition of the states is very
small. We see from Fig. 2 that there exist states with static
moments of order 0.2 with energies only 0.0002$t$ per site above the ground
state. 
In a doped system with some form of magnetic order, the behavior
of the holes in each of the states with different spin
orientations would be identical.
It is reasonable to assume that the effect of the
superposition on the hole behavior is slight.

\begin{figure}
\includegraphics[width=6cm]{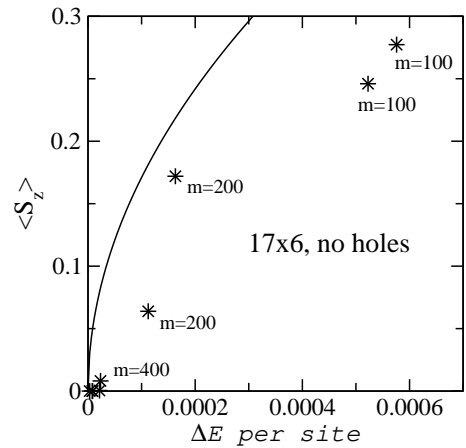}
\caption{Expectation value of the spin $|\langle S_z(l)\rangle|$
at a site in the center of an undoped $17\times6$ $t-J$ system,
as a function of the error in the total energy per site.
The error in the total energy per site is measured relative to a nearly
fully converged DMRG calculation, accurate to about $10^{-6}$. 
Each symbol corresponds to a particular DMRG sweep, for which the number of
states kept was $m$. Two sweeps were performed for each value of $m$.
Although the exact ground state has $S_z(l)=0$, states with substantial
static antiferromagnetic order can have energies only slightly above the
ground state.
The solid curve was generated by applying a staggered magnetic
field to the system, and represents the maximum value of
$|\langle S_z(l)\rangle|$ possible for a given error in the
energy (see text).
}
\label{figtwo}
\end{figure}

We can also analyze this effect in the undoped system by applying a staggered magnetic
field. We assume the energy per site of the system, for small
applied fields,  varies as
\begin{equation}
E(s,h) = E_0 + a s^2 - sh
\label{esh}
\end{equation}
where $h$ is the magnitude of the applied staggered field in the $z$ direction
and $s$ is the average
magnitude of $\langle S_z(l)\rangle$ in response. We minimize $E(s,h)$ over
$s$, keeping $h$ fixed, to find $s(h) = h/(2a)$, and hence $E(s) = E_0 -as^2$.
From several very accurate DMRG simulations with various values of $h$, 
we find  that for $s < 0.1-0.2$, this energy dependence is
accurate, and for the 17$\times$6 system $a\approx0.0034$. 
To describe the case where there is no applied field, and a finite
value of $s$ is considered an error, we define $\Delta E = a s^2$,
yielding $s(\Delta E) = (\Delta E/a)^{1/2}$, which is shown as
the solid line in Fig. 2.
This should be considered
to be an upper bound on the $s(\Delta E)$ obtained from DMRG, for which
there are other sources of error in the wavefunction besides a
finite value of $s$.

\begin{figure}
\includegraphics[width=7cm]{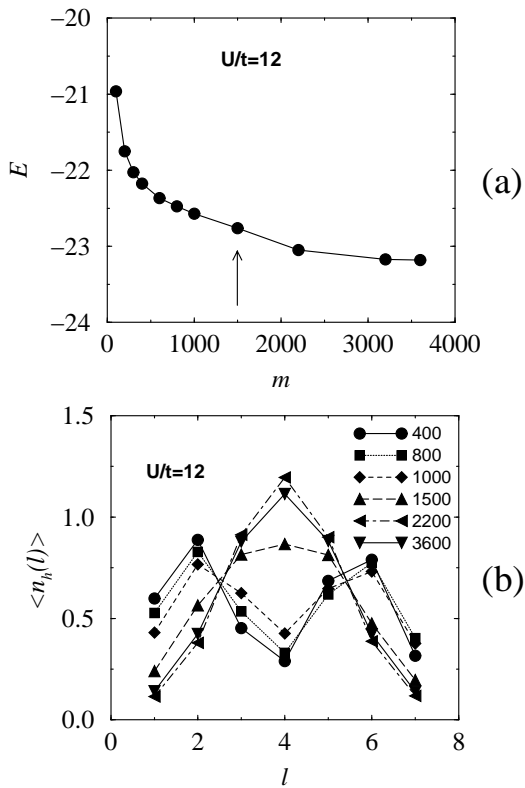}
\caption{(a) The ground state energy of a 7$\times$6 Hubbard model with
U/t $=$ 12 and 4 holes versus the number of basis states kept in the DMRG
calculation. The arrow indicates the approximate point in which the stripe spontaneously
forms. (b) The charge distribution $\langle n_h(l)\rangle$ seen in the DMRG
calculation, labeled by the number of states kept per
block $m$.  } 
\label{figthree}
\end{figure}

We turn now to the Hubbard model. 
Our results for the Hubbard systems were performed using a new
``single-site'' DMRG method\cite{singlesite}, which performs
better than the standard DMRG algorithm having two sites in the
center when the number of states per site is more than two or
three. Using this method, we have been able to keep up to 7500
states per block in some cases. Unfortunately, discarded weights
are not informative and we do not report them.
When performing DMRG
calculations on 2D clusters, one must deal with the possibility
that the calculation will get stuck in a metastable state. For
example, a striped state may be lower in energy than a state
with two widely separated pairs, but the calculation may only
be able to tunnel between these configurations when keeping very large
numbers of states per block. In this case, one must repeat the
calculations with constrained initial configurations, and
compare final energies. On the other hand, a calculation is
particularly robust if one sees that it does tunnel between very
different states.
Fig.~3(a) shows a plot of the ground state
energy for a 7$\times$6 lattice with U/t $=$ 12 and 4 holes as a function of
the number of basis states $m$. The
hole density distribution obtained at a number of sweeps, labeled by the
value of $m$, is shown in
Fig.~3(b). The initial configuration consisted of two separate
pairs.  As the number of basis states increases and the ground state
energy converges, one clearly sees the stripe develop, with the
``tunneling'' occuring for about $m=1200$ states.  From Fig. 4 we see that,
just as for the t-J
model, there is a $\pi$-phase shift in the magnetization density across a
stripe.  
One can see that as the
stripe develops, the DMRG ground state energy decreases and that just as
for the t-J model, doped holes on 6-leg Hubbard ladders can form
striped ground state structures for U/t $=$ 12.

\begin{figure}
\includegraphics[width=4cm]{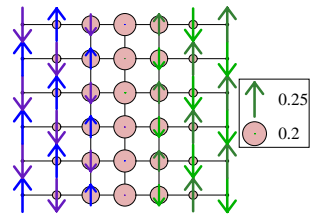}
\caption{The charge and spin distribution for the system of Fig.  3}
\label{figfour}
\end{figure}

\begin{figure}
\includegraphics[width=7cm]{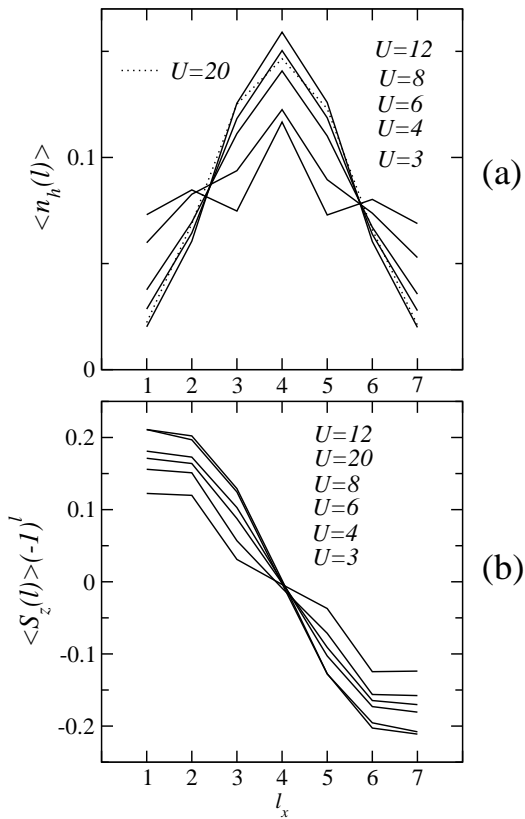}
\caption{(a) The hole density $\langle n_h(l)\rangle$ of a 7$\times$6 Hubbard model 
as a function of the $x$ coordinate $\ell_x$ for various values
of $U$. The solid lines refer to the list of $U$'s on the right,
in order by peak height. The dotted line refers to $U=20$, which
is out of order by peak height.
(b) For the same set of systems the spin density $\langle
S_z(l)\rangle (-1)^{l}$, showing the $\pi$ phase shift of the
stripe. 
} 
\label{figfive}
\end{figure}

We have performed a limited study of the behavior of this system as a function of
U/t.
Fig.~5 shows the charge and spin densities for U/t ranging from
3 to 20.
For U/t$=$8, we started the system with the holes as two
separate pairs. We observed tunneling to the stripe state near
$m=3600$, considerably later than for U/t$=$12.
We let this calculation continue until it used all the
available memory (3 Gb), taking about 1 week of computer time, on an Athlon MP
1800+ processor. In this case, further sweeps reached a maximum of $m=7500$,
with little change in the charge and spin distributions.
For U/t$=$6, we observed tunneling to the stripe state near
$m=4000$, again with a maximum of $m=7500$. For U/t$=$4 and
U/t$=$3, we did not try to observe the tunneling, instead
starting with the four holes together in the center. Here, we
found that a broadened striped configuration was stable, up to the
maximum number of states kept ($m=6000$ and $m=4000$,
respectively). For U/t$=$3, the charge distribution 
is perhaps too broad to consider it a stripe, but the domain
wall nature of the spin configurations configurations is still fairly robust. 
However, since, as noted above, the spin expectation values are
(useful) artifacts of the DMRG procedure, we do not attach much
significance to the specific magnitudes shown in Fig. 5(b).
Fig. 5 also shows results for U/t$=$20. For this run, we started
it in a state with the four holes together, but during the first
several sweeps, keeping only $m \approx 200$, the holes
partially split apart. Subsequently, near $m=1000$, a definite stripe
formed along with the antiferromagnetic domain wall. A maximum
of $m=6000$ states were kept. For this large value of U/t, the
stripe appears somewhat less stable than at U/t$=$12.
We observe a very similar weakening in the stripe in the t-J
model, as the value of J/t is reduced from 0.35 to 0.2 (the peak
height in $\langle n_h(l)\rangle$ drops from 0.21 to 0.18).


In conclusion, we have found that on a moderately sized Hubbard cluster,
with cylindrical boundary conditions and
doped with four holes, the ground state has a stripe. The stripe
is narrow and well-defined for U/t$=$8-12. For smaller values of
U, starting at U/t$\approx$6, the stripe broadens, until at 
U/t$=$3 the width of the stripe is the size of the system. At
U/t$=$20, the stripe is somewhat broadened compared to U/t$=$12.
The overall behavior is very similar to that seen in the t-J model.
Although the open ends in one direction may encourage the stripe
to form, the same roll may be played by neighboring stripes in
larger systems. 
We find nearly the same hole density profile for a stripe in a
7$\times$6 t-J system as in the central portion of a t-J 17$\times$6 t-J
system. For small U/t, the broad stripes we obtain are probably
strongly influenced by the open boundaries.

The would like to that A. Chernyshev for useful comments.
We acknowledge the support of the NSF under grants
DMR98-70930 and DMR98-17242.


\end{document}